\mag=\magstep1

\documentclass[12pt]{article}

\usepackage{latexsym}
\usepackage{amssymb}
\usepackage{amsmath}

\newcommand{\niklabel}[1]{\label{#1}}

\def\{{\lbrace}
\def\}{\rbrace}

\def\cl{{\cal C}\!\ell}

\def\st{\stackrel}

\def\ww{\wedge\ldots\wedge}

\def\R{{\mathbb R}}

\def\F{{\mathbb F}}

\def\E{{\cal E}}

\def\Even{{\rm Even}}
\def\Odd{{\rm Odd}}

\def\be{\begin{equation}}
\def\ee{\end{equation}}

\def\Mat{{\rm Mat}}

\def\be{\begin{equation}}
\def\ee{\end{equation}}

\begin{document}

\title{Generalized exterior algebras}

\author{Nikolay Marchuk}

\maketitle

\thanks{This work is partly supported by the grant NSh-7675.2010.1 and by the Mathematical division of the Russian Academy of Sciences (the programm ``Modern problems of theoretical mathematics'').}

MSC classes: 15A66

\begin{abstract}
Exterior algebras and differential forms are widely used in many
fields of modern mathematics and theoretical physics. In this
paper we define a notion of  $N$-metric exterior algebra, which
depends on $N$ matrices of structure constants. The usual exterior
algebra (Grassmann algebra) can be considered as $0$-metric
exterior algebra. Clifford algebra can be considered as $1$-metric
exterior algebra. $N$-metric exterior algebras for $N\geq2$ can be
considered as generalizations of the Grassmann algebra and
Clifford algebra. Specialists consider models of gravity that
based on a mathematical formalism with two metric tensors. We hope
that the considered in this paper 2-metric exterior algebra can be
useful for development of this model in gravitation theory.
Especially in description of fermions in presence of a gravity
field.
\end{abstract}





\noindent{\bf Clifford algebras with a nondiagonal matrix of structure constants}. Let $\E$ be the $2^n$-dimensional vector space over a field  $\F$ (of real or complex numbers) with basis elements
\be e,e^i,e^{i_1 i_2},\ldots,e^{1\ldots
n},\quad\hbox{где}\quad 1\leq i\leq n,\quad 1\leq i_1<i_2\leq n,
\ldots, \niklabel{cl:basis} \ee
enumerated by ordered multi-indices of length from $0$ to $n$. And let us take a real symmetric nondegenerate matrix $g\in\Mat(n,\R)$ with elements $g^{ij}=g^{ji}$.

For elements of the vector space  $\E$ we define multiplication ({\em Clifford multiplication}) by the following rules:
\medskip

\noindent 1)\label{cl:rules}$\quad(\alpha U)V=U(\alpha
V)=\alpha(UV)$ for $\forall U,V\in {\E}, \alpha\in\F$.

\noindent 2)$\quad(U+V)W=UW+VW,\quad W(U+V)=WU+WV$ for $\forall
U,V,W\in {\E}$.

\noindent 3)$\quad(UV)W=U(VW)$ for $\forall U,V,W\in\E$.

\noindent 4)$\quad eU=Ue=U$ for $\forall U\in {\E}$.

\noindent 5)$\quad e^i e^j+e^j e^i=2g^{ij}e$ for $i,j=1,\ldots,n$.

\noindent 6)$\quad e^{i_1}\ldots e^{i_k}=e^{i_1\ldots i_k}$ for
$1\leq i_1<\cdots<i_k\leq n$.

\medskip
Note that the rules 1)--4) are standard axioms of an associative algebra with identity elements $e$.

Using these 6 rules we can calculate products of any basis elements (\ref{cl:basis}). In fact, to find the product $e^{i_1}\ldots e^{i_k}$ with unordered indices we may rearrange factors in it using the rule 5), taking into account that from the product  $e^i e^j$ in the case $i>j$ we get two summands: one that contain $e^j e^i$ and another that contain $2g^{ij} e$. As a result, with the aid of the rules 1)-5) the product $e^{i_1}\ldots e^{i_k}$ can be transformed into a sum of products of elements $e^i$ with ordered indices
$$
e^{i_1}\ldots e^{i_k}= \alpha e^{l_1}\ldots e^{l_p}+\beta
e^{m_1}\ldots e^{m_q} + \ldots,
$$
where $l_1<\cdots<l_p,\ m_1<\cdots<m_q,\ldots;\
\alpha,\beta,\ldots$ are scalars. Now, in accordance with the rule 6) we obtain
$$
e^{i_1}\ldots e^{i_k}= \alpha e^{l_1\ldots l_p}+\beta e^{m_1\ldots
m_q} + \ldots,
$$
that means at the right hand part we get a linear combination of elements of the basis (\ref{cl:basis}).
Finally, to calculate the product of any two elements of the basis  (\ref{cl:basis}) we must write
$$
e^{i_1\ldots i_k} e^{j_1\ldots j_r}=e^{i_1}\ldots e^{i_k}
e^{j_1}\ldots e^{j_r}
$$
and use the previous reasoning.
\medskip

\noindent{\bf Example}. Let us calculate the product
\begin{eqnarray*}
e^{13}e^{234}&=&e^1 e^3 e^2 e^3 e^4 = e^1(-e^2 e^3+2g^{23}e)e^3 e^4\\
&=&-g^{33}e^1 e^2 e^4+2g^{23}e^1 e^3
e^4=-g^{33}e^{124}+2g^{23}e^{134}.
\end{eqnarray*}

Denote by $E^*$ the vector space that spanned on basis elements  $e^1,\ldots,e^n$ with one index (we use notation $E^*$ because in many applications the space $E^*$ is the dual space to some initial pseudo-Euclidean space  $E$). In the sequel we see that  $2^n$ dimensional vector space $\E$ can be considered as the exterior algebra of the space $E^*$.

The vector space $\E$ over a field $\F$ with the defined operation of multiplication is called (real or complex, depends on $\F$) {\em Clifford algebra} with the matrix of structure constants $g=\|g^{ij}\|$ and denoted by $\cl(E^*,g)$. Basis elements with one index $e^1,\ldots,e^n$ are called {\em generators} of the Clifford algebra  $\cl(E^*,g)$ and the basis (\ref{cl:basis})  is called {\em the Clifford basis} of the algebra $\cl(E^*,g)$.

If the matrix $\|g^{ij}\|$ is diagonal, moreover with $r$ pieces of $1$ and $q$ pieces of $-1$ on the diagonal, then the corresponding Clifford algebra is denoted by  $\cl(r,s)$ (by  $\cl(n)$ for $s=0$).
\medskip

One of the most important properties of Clifford algebras is the following: if an element $U\in\cl(E^*,g)$ is a linear combination of generators  $U=u_j e^j$, then $U^2=(g^{ij}u_i u_j)e$, i.e. the square of this element is a scalar (proportional to the identity element $e$).

\medskip

\noindent{\bf An exterior multiplication of elements of a Clifford algebra $\cl(E^*,g)$}.
For elements of a Clifford algebra $\cl(E^*,g)$ we define the operation of exterior multiplication, which we denote by the symbol  $\wedge$. For products of generators we put
\be e^{i_1}\wedge e^{i_2}\wedge\ldots\wedge
e^{i_k}=e^{[i_1}e^{i_2}\ldots e^{i_k]}, \niklabel{A2} \ee
where square brackets  denote the operation of alternation of indices.

In particular, from  (\ref{A2}) we get the main identity of the Grassmann algebra
$$
e^i\wedge e^j=-e^j\wedge e^i.
$$
\medskip
\noindent{\bf Example.} \be e^{i_1}\wedge
e^{i_2}=\frac{1}{2}(e^{i_1}e^{i_2}-e^{i_2}e^{i_1})=
e^{i_1}e^{i_2}-g^{i_1 i_2}e, \niklabel{N2} \ee

It can be checked that the formula (\ref{A2}) is equivalent to the following formula:
\be e^{i_1}\wedge\ldots\wedge e^{i_k}=e^{i_1}\ldots
e^{i_k}+ \sum_{r=1}^{[\frac{k}{2}]} \frac{(-1)^r}{r!}
Q^r(e^{i_1}\ldots e^{i_k}), \niklabel{A6} \ee
where
$$
Q(e^{i_1}\ldots e^{i_k})= \sum_{1\leq p<q\leq k} (-1)^{q-p-1}
g^{i_p i_q}e^{i_1}\ldots\check{e^{i_p}}\ldots\check{e^{i_q}}\ldots
e^{i_k}.
$$
The symbol $\check{}$ over the factor $\check{e^{i_p}}$ means that this factor in the product is omitted,
$Q^r$ is the operation $Q$ applied $r$ times, and  $[\frac{k}{2}]$ is the integer part of the number $k/2$.

Formula (\ref{A6}) can be taken as the definition of exterior multiplication of Clifford algebra elements instead of formula (\ref{A2}).

Using formula (\ref{A6}) we may express Clifford products of generators $e^i$ in terms of exterior products of generators. Namely,
\be e^{i_1}\ldots e^{i_k}=e^{i_1}\wedge \ldots\wedge
e^{i_k}+ \sum_{r=1}^{[\frac{k}{2}]} \frac{1}{r!}
Q^r(e^{i_1}\wedge\ldots\wedge e^{i_k}), \niklabel{A7} \ee
where
$$
Q(e^{i_1}\wedge\ldots\wedge e^{i_k})= \sum_{1\leq p<q\leq k}
(-1)^{q-p-1} g^{i_p i_q}e^{i_1}\wedge\ldots\wedge
\check{e^{i_p}}\wedge\ldots\wedge\check{e^{i_q}}\wedge\ldots\wedge
e^{i_k}.
$$
\medskip

\noindent{\bf Example.} From formula (\ref{A7}), in particular, we get
\be
e^{i_1}e^{i_2}=e^{i_1}\wedge e^{i_2}+g^{i_1 i_2}e, \niklabel{A8}
\ee
$$
e^{i_1}e^{i_2}e^{i_3}=e^{i_1}\wedge e^{i_2}\wedge e^{i_3} +g^{i_2
i_3}e^{i_1}-g^{i_1 i_3}e^{i_2}+g^{i_1 i_2}e^{i_3},
$$
Formula (\ref{A7}) gives us possibility to express elements of Clifford basis (\ref{cl:basis}) in terms of linear combinations of the following elements, which form a new basis of Clifford algebra (Grassmann basis)
\be e,e^i,e^{i_1}\wedge
e^{i_2},\ldots,e^{1}\wedge\ldots\wedge e^n, \quad\hbox{где}\quad
1\leq i\leq n,\quad 1\leq i_1<i_2\leq n, \ldots.
\niklabel{gr:basis} \ee
Conversely, formula  (\ref{A6}) gives us possibility to express elements of Grassman basis (\ref{gr:basis}) in terms of linear combinations of elements of Clifford basis  (\ref{cl:basis}).

Now we can find the result of exterior multiplication
\be
e^{i_1\ldots i_p}\wedge e^{j_1\ldots j_q} \niklabel{A12} \ee
of any elements of Clifford basis. This can be done it tree steps.
\medskip

G1. Let us express basis elements $e^{i_1\ldots i_p},e^{j_1\ldots
j_q}$ in terms of elements of Grassmann basis (\ref{gr:basis}) and substitute corresponding expressions into (\ref{A12}).

G2. Further we calculate the exterior product of elements of Grassmann basis. We get a result in the form of a sum of basis elements (\ref{gr:basis}).

G3. Using formulas (\ref{A6}), we write down result in terms of Clifford basis (\ref{cl:basis}).
\medskip

\noindent{\bf Example.}
\begin{eqnarray*}
e^{13}\wedge e^{23}&=&(e^1\wedge e^3+g^{13}e)\wedge(e^2\wedge e^3+g^{23}e)\\
&=&g^{23}e^1\wedge e^3+g^{13}e^2\wedge e^3+g^{13}g^{23}e\\
&=&g^{23}(e^{13}-g^{13}e)+g^{13}(e^{23}-g^{23}e)+g^{13}g^{23}e\\
&=&g^{23}e^{13}+g^{13}e^{23}-g^{13}g^{23}e.
\end{eqnarray*}

On the same way we may find the result of Clifford multiplication of any elements of Grassmann basis
\be (e^{i_1}\wedge\ldots\wedge
e^{i_p})(e^{j_1}\wedge\ldots\wedge e^{j_q}), \quad
i_1<\cdots<i_p;\ j_1<\cdots<j_q. \niklabel{A13} \ee
This also can be done in tree steps.
\medskip

C1.\label{C1C2C3} Let us express Grassmann basis elements $e^{i_1}\wedge\ldots\wedge e^{i_p}$, $e^{j_1}\wedge\ldots\wedge e^{j_q}$ in terms of elements of Clifford basis (\ref{cl:basis}) using formulas (\ref{A6}) and substitute corresponding expressions into (\ref{A13}).

C2.  Further we calculate the Clifford product of elements of Clifford basis. We get a result in the form of a sum of basis elements (\ref{cl:basis}).

C3. Using formulas (\ref{A7}), we write down result in terms of Grassmann basis (\ref{gr:basis}).
\medskip

\noindent{\bf Example.} It is easy to check that
$$
(e^1\wedge e^3)(e^2\wedge e^3)=-g^{33}e^1\wedge
e^2+g^{23}e^1\wedge e^3- g^{13}e^2\wedge
e^3+(g^{13}g^{23}-g^{12}g^{33})e.
$$

Therefore we arrive at the $2^n$-dimensional vector space over a field $\F$ (real or complex numbers) with two operations of multiplication -- Clifford multiplication and exterior multiplication and with two bases  (\ref{cl:basis}) and (\ref{gr:basis}). For both operations of multiplication the associativity and distributivity axioms are satisfied. The basis element $e$ is the identity element for both operations.

If the matrix of structure constants $g=\|g^{ij}\|$ is diagonal, then every formulas (\ref{A6}) and (\ref{A7}) gives us relations
\be e^{i_1}\ldots
e^{i_k}=e^{i_1}\wedge\ldots\wedge e^{i_k}\quad \hbox{при}\quad
i_1<\cdots<i_k \niklabel{A15} \ee
that means bases (\ref{cl:basis}) and (\ref{gr:basis}) are coincide.
In this simple case the described construction of  $2^n$-dimensional
 vector space with two operations of multiplication
 (exterior and Clifford) was considered by many authors
 (the first was H.~Grassmann in 1877 \cite{grassmann},
 see also  \cite{doran}).

\medskip

\noindent{\bf Exterior polymetric algebras}. Clifford algebra $\cl(E^*,g)$, considered with only operation of exterior multiplication $\wedge$ and with Grassmann basis (\ref{gr:basis}), is the exterior algebra and denoted by $\Lambda(E^*)$. Hence in previous section we start from Clifford algebra $\cl(E^*,g)$ and arrive at exterior algebra $\Lambda(E^*)$. In this section we follow in the opposite direction -- from exterior algebra $\Lambda(E^*)$ to Clifford algebra $\cl(E^*,g)$. On this way we arrive at a new class of mathematical objects -- {\em exterior polymetric algebras}.
\medskip

Let  $E^*$ be an $n$-dimensional vector space with the basis $e^1,\ldots,e^n$  and $\Lambda(E^*)$ be the exterior algebra of the vector space $E^*$ with the operation of exterior multiplication $\wedge :
\Lambda(E^*)\times \Lambda(E^*)\to \Lambda(E^*)$ and with Grassmann basis  (\ref{gr:basis}).

The exterior multiplication satisfies the condition $e^i\wedge e^j=-e^j\wedge e^i$. The exterior algebra $\Lambda(E^*)$ is an associative algebra with identity elements $e$. The dimension of the exterior algebra is equal to  $2^n$.
\medskip

Let us take a symmetric nondegenerate matrix
$g=\|g^{ij}\|\in\Mat(n,\R)$. With the aid of this matrix and using
formulas (\ref{A7}) we define new basis (\ref{cl:basis}) of the
algebra $\Lambda(E^*)$. This basis (Clifford basis) gives us
possibility to define Clifford multiplication for elements of the
exterior algebra using rules 1)-6) (see page \pageref{cl:rules})
and rules C1-C3 (see page \pageref{C1C2C3}).

So we again arrive at the algebra $\cl(E^*,g)$ with two operations
of multiplication -- exterior and Clifford. However the fact that
now we start from the exterior algebra gives us possibility to
generalize the considered construction. Namely, we may consider
algebras that have several matrices of structure constants
$g_1,\ldots,g_N$.

 That is let $E^*$ be $n$ dimensional vector space and
 $\Lambda(E^*)$ be the exterior algebra of the space $E^*$ with
 operation of exterior multiplication (the dimension of vector
 space $\Lambda(E^*)$ is equal to $2^n$) and with the basis (\ref{gr:basis}).
 And let we have $N$ symmetric nondegenerate matrices $g_1,\ldots,g_N\in\Mat(n,\R)$.
Elements of these matrices we denote by $g^{ij}_{(k)}$,
$i,j=1,\ldots,n$, $k=1,\ldots,N$. With the aid of matrices
$g_1,\ldots,g_N$ we define $N$ operations of multiplication
$\st{k}{\vee} : \Lambda(E^*)\times\Lambda(E^*)\to\Lambda(E^*)$,
$k=1,\ldots,N$. Every operation $\st{k}{\vee}$ is an operation of
Clifford multiplication of elements of exterior algebra that
defined with the aid of the matrix $g_k$ and using rules 1)-6)
(see page \pageref{cl:rules}) and rules C1-C3 (see page
\pageref{C1C2C3}).

The resulting unital associative algebra (with the exterior
multiplication $\wedge$ and with Clifford multiplications
$\st{1}{\vee},\ldots,\st{N}{\vee}$) is called {\em exterior
polymetric algebra or $N$-metric algebra} and denoted by
$\Lambda(E^*,g_1,\ldots,g_N)$. $0$-metric algebra coincide with
the exterior algebra $\Lambda(E^*)$ and $1$-metric algebra
coincide with the Clifford algebra $\cl(E^*,g)$. $N$-metric
algebras $\Lambda(E^*,g_1,\ldots,g_N)$ for $N\geq 2$ can be
considered as a generalization of exterior algebra and Clifford
algebra.
\medskip

In the considered construction of exterior polymetric algebra the
Grassmann basis (\ref{gr:basis}) play a primary role and Clifford
bases play secondary role.

A notion of rang of an element of exterior polymetric algebra
corresponds to the Grassmann basis. Namely, an element
$U\in\Lambda(E^*,g_1,\ldots,g_N)$ of the form
$$
U=\sum_{j_1<\ldots<j_k}u_{j_1\ldots j_k}e^{j_1}\ww e^{j_k}
$$
is called {\em element of rang $k$}. The set of rang $k$ elements
is the subset $\Lambda_k(E^*,g_1,\ldots,g_N)$ of the dimension
$C_n^k$. Also we define notions of even and odd elements of a
polymetric algebra and
\begin{eqnarray*}
\Lambda(E^*,g_1,\ldots,g_N)&=&\oplus^n_{k=0}\Lambda_k(E^*,g_1,\ldots,g_N)\\
&=&\Lambda_\Even(E^*,g_1,\ldots,g_N)\oplus\Lambda_\Odd(E^*,g_1,\ldots,g_N).
\end{eqnarray*}
Dimensions of subspaces $\Lambda_\Even(E^*,g_1,\ldots,g_N)$,
$\Lambda_\Odd(E^*,g_1,\ldots,g_N)$ are equal to $2^{n-1}$. The set
of  even elements $\Lambda_\Even(E^*,g_1,\ldots,g_N)$ is the
subalgebra of the exterior $N$-metric algebra.

Operations of conjugation of elements of a polymetric algebra also
correspond to the Grassmann basis. In particular, the operation of
pseudo-Hermitian conjugation $\ddagger
:\Lambda(E^*,g_1,\ldots,g_N)\to\Lambda(E^*,g_1,\ldots,g_N)$ can be
defined by the following rules:
\begin{itemize}
\item $(e^a)^\ddagger =e^a$, $a=1,\ldots,n$;

\item $(U\wedge V)^\ddagger =V^\ddagger \wedge U^\ddagger $;

\item $(U+V)^\ddagger =U^\ddagger +V^\ddagger $;

\item $(\lambda U)^\ddagger =\bar\lambda U^\ddagger $.
\end{itemize}
где $U,V\in\Lambda(E^*,g_1,\ldots,g_N)$, $\lambda\in\F$.

\medskip

About applications of the polymetric exterior algebra. In addition
to Einstein's General Theory of Relativity there are several
alternative models of gravity. In particular, specialists consider
models that based on a mathematical formalism with two metric
tensors. We hope that the considered in this paper 2-metric
exterior algebra can be useful for development of this models in a
gravitation theory. Especially for description of fermions in
presence of a gravity field.


\newpage

Steklov Mathematical Institute RAS

Marchuk Nikolay Gur'evich

Gubkina str.8, Moscow 119991, Russia

Phones

Office: +7-499-1351449


e-mails: nmarchuk@mi.ras.ru, nmarchuk2005@yandex.ru


\begin{thebibliography}{99}

\bibitem{mybook} Marchuk N.G., {\sl Field theory equations and Clifford
algebras} (in Russian), Izevsk, R\&C Dymamics, (2009).\par

\bibitem{grassmann} Grassmann H., Math. Commun. 12, 375 (1877).

\bibitem{doran} Doran C., Hestenes D., Sommen F., Van Acker N.,  J.Math.Phys. 34(8),
(1993) 3642.\par

\bibitem{benn} Benn I.M., Tucker R.W., {\sl An introduction
to spinors and geometry with applications to physics}, Bristol,
(1987).\par

\bibitem{lounesto} Lounesto P., {\sl Clifford Algebras and Spinors},
Cambridge Univ. Press (1997, 2001)\par

\end{thebibliography}
\end{document}